# Controlling electromagnetic scattering with wire metamaterial resonators

DMITRY S. FILONOV[1, 2,*], ALEXANDER S. SHALIN[1], IVAN IORSH[1], PAVEL A. BELOV[1], AND PAVEL GINZBURG[2]

[1]*ITMO University, St. Petersburg 197101, Russia*
[2]*School of Electrical Engineering, Tel Aviv University, 69978, Israel*
*Corresponding author: *dmitriif@mail.tau.ac.il

**Manipulation of radiation is required for enabling a span of electromagnetic applications. Since properties of antennas and scatterers are very sensitive to a surrounding environment, macroscopic artificially created materials are good candidates for shaping their characteristics. In particular, metamaterials enable controlling both dispersion and density of electromagnetic states, available for scattering from an object. As the result, properly designed electromagnetic environment could govern waves' phenomena. Here electromagnetic properties of scattering dipoles, situated inside a wire medium (metamaterial) are analyzed both numerically and experimentally. Impact of the metamaterial geometry, dipole arrangement inside the medium, and frequency of the incident radiation on scattering phenomena was studied. It was shown that the resonance of the dipole hybridizes with Fabry–Pérot modes of the metamaterial, giving rise to a complete reshaping of electromagnetic properties. Regimes of controlled scattering suppression and super-scattering were observed. Numerical analysis is in an agreement with experiments, performed at the GHz spectral range. The reported approach to scattering control with metamaterials could be directly mapped into optical and infrared spectral ranges by employing scalability properties of Maxwell's equations.**

## 1. INTRODUCTION

Radiation and scattering properties of either optical emitters or classical antennas, embedded in a medium, could be substantially different from those of isolated structures. These effects are quite commonly observed with radio waves, where couplings, cross-talks, and other concomitant effects should be taken into account for a complete analysis or engineering of certain characteristics. For example, performances of directional antennas rely on careful design of interference between a feeding element and a collection of resonant scatterers (impedance matching) [1]. Similar effects in the optical domain are usually treated via photonic density of states concept; here the spontaneous emission rate enhancement with a photonic cavity is referred by the name of Purcell effect [2]. However, radiation efficiencies at any spectral range could be described with a unified mathematical approach, based on classical electromagnetic Green's functions [1]. From the classical point of view, this result is the direct consequence of the Maxwell's equations scalability in respect to an operation frequency. Remarkably, quantum dynamics of emission processes in both weak and strong coupling regimes could be analyzed with the help of classical electromagnetic Green's function, as a result of the fundamental causality and fluctuation-dissipation relations [3]. This classical-quantum correspondence enables experimental investigations and emulations of complex photonic processes with lower frequencies, where both fabrication and measurements are routinely available. However, the distinctive discrepancy between electromagnetic phenomena at different frequencies is the lack of scalability in material parameters, since complex permittivities and permeabilities are frequency-dependent. These properties, being pre-determined by the nature, could be, however, modified by subwavelength structuring. This concept of metamaterials (e.g. [4], [5]), has been successfully employed for obtaining intriguing material properties, such as negative ε and μ [6] with following demonstrations of invisibility cloaking [7] and other remarkable phenomena. Impact of structured environment on radiation properties of quantum emitters gained additional attention with introducing the concept of "hyperbolic metamaterials" [8]. The key property of the hyperbolic media is to deliver extremely high and broadband Purcell factors, virtually limited by granularity of structures' realization [9]. Additional feature of this macroscopic type of enhancement, is its relative insensitivity to emitter's position within the structure, which is in direct contrast to the antenna approach, relying on localized fields and interference phenomena [10], [11].

While the impact of metamaterials on radiation efficiencies is widely studied, their ability to manipulate electromagnetic scattering properties of embedded objects is less explored. Nevertheless, both of these phenomena are tightly related. Both could be described with either density of electromagnetic states concept or by utilizing electromagnetic Green's functions approach. However, elastic versus inelastic aspects of interactions should be properly addressed. As a additional example, discrete diffraction phenomenon in waveguide arrays [12] utilizes controllably reduced number of electromagnetic modes, available for an interaction. Metamaterials exploit similar approach of configuring spatial dependence of modes with additional remarkable flexibility of controlling their density of states.

Here comprehensive numerical and experimental investigations of electromagnetic scattering from dipoles, embedded in a finite-size wire medium are reported. Fundamental scattering properties were investigated by varying dipoles' lengths (90-110 mm), their spatial arrangements, and radiation frequencies (in the range of 0.2-2 GHz). Special attention was paid to whether dipoles have or have not a direct electrical contact with the wires, forming the metamaterial (Fig. 1). The performed analysis paves the way to addressing a variety of fundamental light-matter interaction phenomena by means of emulation experiments, as will be discussed hereafter.

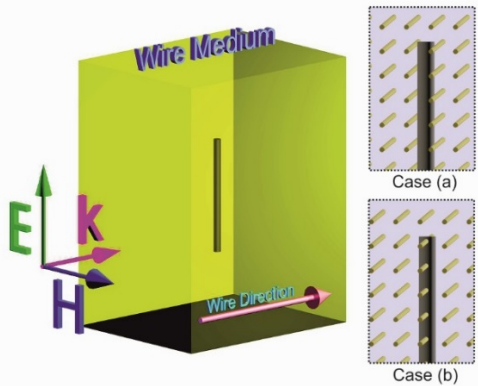

**Fig. 1.** (Color online) Schematics of the considered structure: finite sized metamaterial (wire medium) with an embedded scatterer (a dipole). Two interaction scenarios are considered - (a) noncontact mode - dipole is not touching the wires and (b) contact mode – dipole is touching the wires.

## 2. RESULTS

Media, composed from parallelly-aligned perfectly conducting wires, have many remarkable electromagnetic properties [13]. These densely packed wires prohibit the propagation of electromagnetic waves polarized along their direction and support TEM type of modes (both electric and magnetic fields are perpendicular to the propagation direction). This phenomenon results in degenerately flat dispersion diagrams, and the subwavelength (yet finite) periodicity leads to inherently strong spatial dispersion [14]. Field distributions of modes, supported by the structure, are substantially different from those of the free space and have remarkable properties that could be employed for various applications, e.g. imaging beyond the classical (free space) diffraction limit (see ,e.g., [15]). It is worth noting, that the optical counterpart of densely packed nanorods-array could support elliptical, hyperbolic and compound (nonlocal) dispersion regimes, depending on a frequency [16], [17]. This phenomenon could be also employed for sub-diffraction 'hyperlensing' [18] and other applications. In all of the above cases (optical and microwave) the density and spatial distribution of electromagnetic modes are substantially different from those of the free space and, as a result, severe impact on scattering processes and diffraction is expected. Those effects will be underlined, during discussion of results, presented in the following sections.

### 2.1 Dipoles in a wire medium – a noncontact mode

In order to obtain efficient scattering from a perfectly conducting object, the physical dimensions of the latter should be comparable with the excitation wavelength. It is worth noting, that in the optical domain subwavelength objects could have enormously large scattering cross-sections by exploiting localized plasmon resonance phenomenon [19],[20]. Scattering from thin metal wires in the microwave regime has the first resonance, when object's length along the polarization direction of the incident wave reaches the half-wavelength scale. Hereafter, dipolar antennas (will be called dipoles here) with lengths of 90, 100 and 110 mm, and 8 mm in diameter will be considered. Total scattering (radar) cross-sections (RCS) of the dipoles in free space appear on Fig. 2(a), where different color lines correspond to three different dipole lengths. All the numerical results were obtained with the help of finite element method [21] (additional information appears in 'Methods' section). At the next step those half-wavelength (at different frequencies) dipoles were situated inside the wire medium *without* having a direct electrical contact with the surrounding structure (Fig. 1, inset (a)). The metamaterial consists of 16x26 array of thin metal wires, with diameters of 2mm. The dipole orientation is perpendicular to the direction of the wires, as appears on Fig. 1 - insets. The resulting RCSs appear on Fig. 2(b). The RCSs' spectra of dipoles in the wire medium are red-shifted with respect to the free space scenario, and peaks fall within the range of 0.7-1 GHz (major contribution of the wires). Furthermore, the set of blue-shifted resonances (1.5-1.8 GHz) and the collection of fast-oscillating features appear on the spectra. The later results from the contribution of the higher order modes of the finite size wires array and will be the subject of a future investigation. The main focus will be made on the strong interaction between the scatterer and the low-order modes of the metamaterial structure.

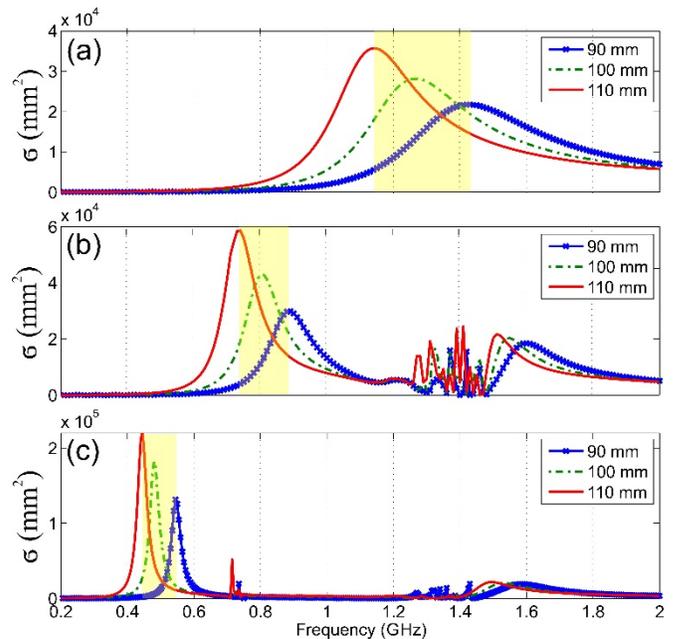

**Fig. 2**. (Color online) Scattering/radar cross-sections of dipoles in (a) free space (b) in wire medium, noncontact mode (c) in wire medium, contact mode. Solid blue with dots, dash-dotted green, and solid red lines correspond to the dipole lengths of 90, 100, and 110 mm. The metamaterial consists of 16x26 (scatterer does not intersect with the wires) and 17x26 (scatterer intersects with the wires) thin metal wires, with 2 mm diameter and 200 mm length.

Since the scattering dipole is situated at the point of high symmetry of the metamaterial (at the center), it could couple only to the high symmetry modes - even in this case. The major RCS peak corresponds to the lowest even mode, while the second one (at the higher frequency) is the signature of the interaction with the next second symmetric mode. For the extended frequency range, the signatures of higher order modes will be observed. Since the excitation of the odd modes is symmetry-forbidden, there is a scattering suppression at the entire gap in between two major RCS peaks. This type of scattering suppression in the vicinity of the super-scattering spectral window is quite common. For example, consideration of fundamental causality and energy conservation principles in analysis of frequency-dependent scattering properties leads to an appearance of strong extinction peaks in the spectral vicinity of suppressed areas due to existence of certain sum rules [22], unless diamagnetic materials are involved [23]. The spectral positions of the RCS peaks on Fig. 2(b) could be tailored by varying parameters of the metamaterial – specifically the wires' length.

Fig. 3(a) shows the RCS dependence for the set of six wires' lengths (50 - 300 mm) in the case of the noncontact mode (scattering dipole is not touching the wires), while the dipole's dimensions were kept the same (110 mm). The increase of the wire length shifts the first Fabry–Pérot resonance to lower frequencies, as could be seen in Fig. 3(a). Similar behavior takes place in classical Fabry–Pérot etalons, where increase of the bulk material thickness leads to the red shift of the first fundamental resonance. The behavior of the second even mode is similar - it replicates the trends of the main one. It should be noted, however, that the shifts do scale linearly with changing the wires length, but the proportionality coefficient is not unity, as could be expected from a simple resonator theory. Nontrivial scaling factor in finite size metamaterial cavities results from both waves dispersion inside the metamaterial and other two dimensions that were kept constant [24].

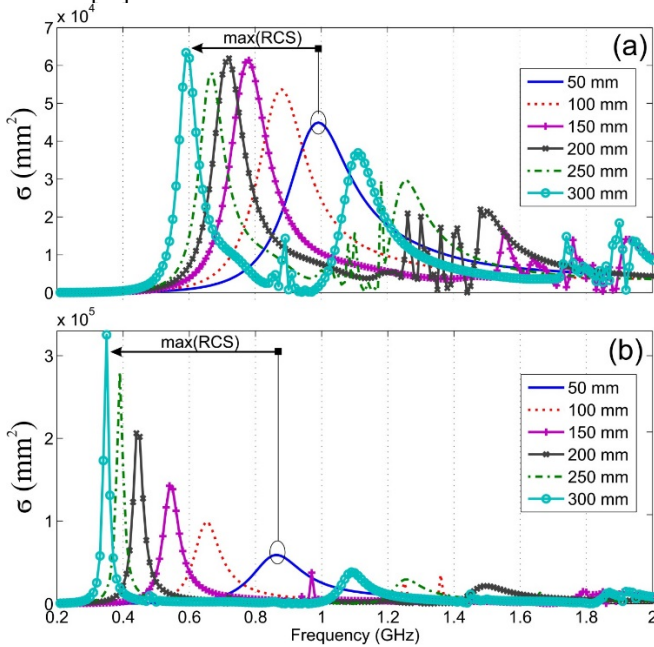

**Fig. 3**. (Color online) Scattering/radar cross-sections of 110 mm dipole in metamaterial resonators with for different length of constituting wires (a) noncontact mode (b) contact mode. The length of the wires changes from 50 to 300 mm (from blue to cyan) in steps of 50 mm. Other parameters of the structure are the same as those, used for Fig.2.

## 2.2 Dipoles in a wire medium – a contact mode

In the next set of investigations the scattering dipole was placed in the direct contact with the wires. In order to preserve the symmetry of the structure, the metamaterial consists of 17x26 wires (dipole touches the 9th wire). The introduction of this electrical contact changes the nature of the coupling phenomenon from the capacitive (noncontact mode) to the conduction. Metamaterial's rods now experience a shortage at their middle, enabling a free flow of current between few connected wires. The RCSs for three different dipoles intersecting the middle row of metamaterial wires are shown in Fig. 2(c). The distinctive difference between these results and the noncontact case is the stronger red shift of the main resonance and its narrowing. The red shift results from effectively longer wire, moreover, the u-band shape of shortened conductors leads to concave geometry, also responsible for the red shift [25]. It should be mentioned that in the contact case the field is more localized within the wire medium which is the reason for Q-factor increase and the corresponding resonances narrowing. The fields' structure will be discussed in details at the next section. The RCS's peaks are strongly dependent on the length of wires, constituting the metamaterial, similarly to the noncontact case. Fig. 3(b) shows the structure of scattering peaks for different geometries. Much more pronounced and broadband scattering suppression window is observed in the contact mode case, compared to the noncontact scenario. The main reason for this occurrence is the vanishing current on the dipolar scatterer, as it is physically separated to a collection of small sections - each one shortened by the metamaterial's wires. Distinctive difference between contact and noncontact modes originates from allowing/disallowing the current flow along the scattering dipole. This intuitive description will be supported by numerical results, represented at the next section.

Similar phenomena occur in optics, where capacitive or conductive couplings between non-touching/touching noble metal particles control the spectral response of localized plasmon resonances [25]. Moreover, the narrowing of the resonance (quality factor increase) is also an attribute of plasmonic resonances [25]. The difference between optical and GHz spectral ranges lies in the nature of the involved electromagnetic currents – in optics polarization currents should be considered instead of the conduction ones at lower frequencies.

## 2.3 Electromagnetic fields inside the wire medium

Structure of electromagnetic fields inside the metamaterial resonator will be investigated next in order to verify behaviors at scattering suppression and super-scattering regions. Figs. 4(a- d) show the intensity distribution of electric fields in the systems, considered in Fig. 3. Fig. 4(a) shows the field intensity at the frequency of 1.12 GHz, where the RCS of the 110 mm dipole has maximum in the free space as shown in Fig. 2(a), while the plots for noncontact and contact configurations at the same illumination frequency appears in Figs. 4 (b) and (d), respectively. Crossections are given for the back facet (x=0), the middle plane including the dipole (y=0), and the bottom side (z=100mm) of the metamaterial parallelepiped. Plots in the right column of Fig. 5 correspond to the frequency of the major RCS maximum, affected by the metamaterial – noncontact and contact modes in panels (c) and (e) respectively. Frequencies of the maxima are 0.74 and 0.445 GHz.

The scattering suppression is observed for both noncontact and contact cased, as could be observed from both RCSs values Fig. 2 (b) and (c) and the field distributions Fig. 4 (b) and (d). While in both cases flat wave fronts are relatively undistorted by the dipolar scatterer, the contact case delivers better scattering suppression, as the effective shortage of the dipole with wires, composing the metamaterial, obstacles the current flow along the dipole as previously stated. Field intensify it the dipole's position is minimal and approaches zero, resulting in vanishing current on the conduction due to associated boundary conditions.

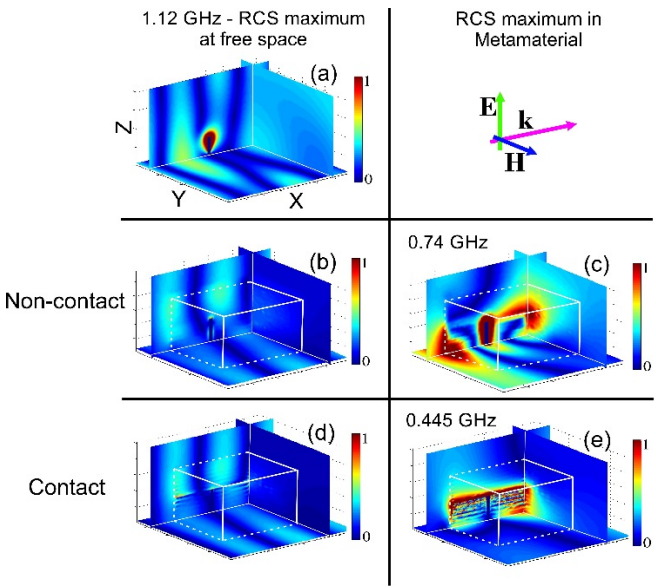

**Fig. 4.** (Color online) Field intensity maps (a, b, d - 1.12 GHz, c - 0.74 GHz, e - 0.445 GHz). (a) – dipole in a free space, (b, c) – noncontact mode, (d, e) – contact mode. The structure is excited with a plane wave, propagating from the left to the right (upper-left panel). Other parameters of the structure are the same as those used for Fig. 2.

Strong interaction between the dipolar scatterer and the metamaterial causes the reshaping of the RCS structure. In particular, as it was already mentioned above, the major RCS peaks are red shifted. The fields at those new RCSs maxima appear on Fig. 4 (c) and (e). Fields at 0.74 GHz in the noncontact mode are strongly enhanced by the entire structure, which collectively resonates with the incident field. Well defined field maximum is obtained on the dipole, which impedance is better match to the free space with the help of metamaterial resonator. On the other side, in the contact mode at 0.443 GHz the field on the dipole vanishes due to the beforehand mentioned shortage. The main contribution to the radiation is delivered now by the metamaterial. As it could be seen, metamaterial wires act as a collection of antennas, reradiating the excitation. All those field behaviors are consistent with the far field signatures, obtained via RCS calculations (Fig. 2).

## 2.4 Dipoles, displaced from the center of the wire medium

The systems, considered above, have distinctive reflection symmetry. This symmetry prohibits the excitation of odd modes, which could emerge in the spectrum once the scatterer is displaced from the metamaterial's center. Figs. 4 (a, b) show the changes in spectra for noncontact and contact modes in cases in which the scatterer has been shifted from the symmetry point. Appearance of odd modes could be clearly seen in both cases. As the modes have finite spectral width, they overlap and interfere with each other. This effect is pronounced for the noncontact case (Fig. 4(a)) and manifests itself in fast RCS's oscillations in the vicinity of the major peak – 0.7GHz. In the contact case, this effect is less important, nevertheless the odd modes affect the scattering suppression region. The effective bandwidth of the frequency range of scattering suppression is reduced and odd modes peaks appear within the gap.

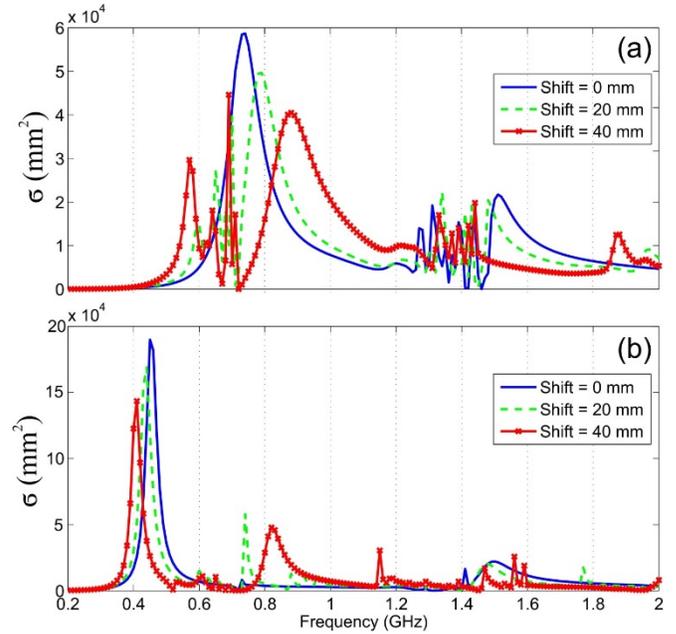

**Fig. 5.** (Color online) Scattering/radar cross-sections of 110 mm dipole in metamaterial, made of 200 mm length wires (a) noncontact mode (b) contact mode. 3 curves on each panel correspond to the dipole position in respect to the structure's centre. Plots corresponding to 0, 20, and 40 mm shifts marked by solid blue, dashed green and solid red with dots. The parameters of the wire medium are the same as those, used for Fig.2.

## 2.5 Experimental observation of scattering on a dipole in the noncontact mode

In order to verify the numerical predictions, experimental studies were performed in an anechoic chamber. The measurements were conducted in the following manner: a rectangular horn antenna (TRIM 0.75 GHz to 18 GHz; DR) was connected to a transmitting port of the vector network analyzer Agilent E8362C. This configuration replicates a plane wave excitation with a good accuracy. The wire medium was placed into the far-field region of the antenna and the similar horn (TRIM 0.75 GHz to 18 GHz) was employed as a receiver (3.5 m is the distance between the horns). The medium consists of 16x26 array of parallel brass wires (2 mm diameter and 200 mm length). Period of the structure along X and Y directions is equal to 10 mm. The effective scattering cross-sections were extracted from imaginary parts of the forward scattering amplitudes (according to the optical theorem) [26].

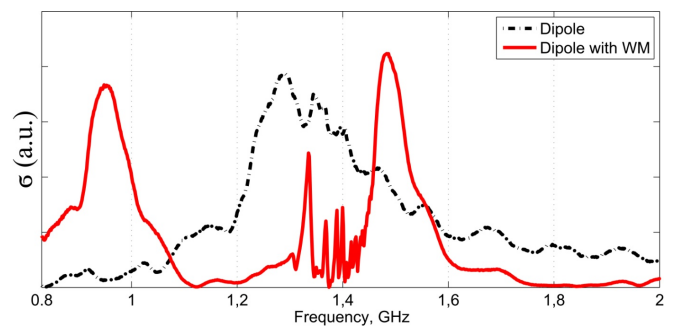

**Fig. 6.** (Color online) Experimental scattering/radar cross-sections (arb. units). Red solid line – 100 mm dipole in 200 mm metamaterial

(noncontact mode), black dot-dashed line – the same dipole in the free space. The parameters of the wire medium are the same as those used for Fig. 2.

Experimentally obtained RCSs appear on Fig. 6 – black dotted line represents the dipole in the free space, while red solid line stays for the scattering inside the metamaterial (noncontact mode). Comparing the experiment with the numerical predictions (Figs. 2(a, b)), enable observing a good correspondence. It should be noted, however, that relative heights of RCS peaks, obtained with the simulation and experiment, do not agree. This effect is generally related to a complexity of a free space setup calibration over a wide range of frequencies. Furthermore, sensitivity of vector network analyzer drops fast with lowering the frequency, making noise impact to be more severe. Additional undesired oscillation behavior in the experimental scattering spectra corresponds to a set of artificial Fabry–Pérot resonances between the wire medium interface and both transmitting and receiving horn antennas. Additional discrepancies between the experimental and numerical results emerge mainly due to the wavelength dependent wave front distortions of the incident wave, produced by the horn antenna and are quite commonly appear in lab experiments of this kind. Nevertheless, strong scattering suppression at GHz wide spectral window, predicted by numerical analysis, was confirmed experimentally.

## 2.6 Details of numerical method

All the numerical results were obtained with the help of finite element method and subsequently verified with a time-domain solver [21]. Finite-size structures were placed in a large simulation domain (far field) with scattering boundary conditions imposed. Variable mesh with overall 250000 elements was utilized and stability of solutions was checked. Plane wave excitation was generated by one port, implemented on the left facet of the simulation domain, while the collecting port on the opposite side enabled calculating forward scattering. Optical theorem was employed for calculating total RCS.

## 3. Discussion and Conclusion

The scalability of physical laws (especially Maxwell's equations) in respect to a dimensionless parameter (in electrodynamics it is wavelength over characteristic dimension of a system) is a very powerful tool. Scaling up systems physical dimensions makes their fabrication and measurements to be more straightforward and gives enormous advantages for detailed investigations. Here by performing scattering experiments with microwaves in structured media we have advanced an emulation tool, enabling investigations of complex light-matter interactions at the nanoscale. In particular, it is quite useful to consider a counterpart of emission/scattering resonance shifts, taking place in the optical domain. Dynamically stimulated shift of quantum energy levels is referred as the AC Stark effect, while the shift due to the virtual photons exchange with the surrounding is referred as the Lamb shift [27]. Observation of those effects is relatively complex, especially once auxiliary nanostructures are involved. However, in both cases (AC Stark and Lamb shifts) classical–quantum correspondence could be drawn and, as a result, all the information is encapsulated in the knowledge of electromagnetic Green's function. For example, the Lamb shift is proportional to the real part of the Green's function [3], nevertheless the virtual photon exchange cannot be mapped directly due to the strong thermal noise above the vacuum limit in the case of lower frequencies. However, in the optical domain the observation of those shifts is much more challenging than with radio waves, due to the naturally small dipolar moments. Nevertheless, general trends in optics could be predicted with RF modeling and experiments, underlining the significance of the emulation experiments. Furthermore, as was recently shown, relatively small (orders of tens) finite arrays of rods could predict the light-matter interaction behavior of infinitely large systems [24], [28]. The experimental platform, discussed at section 2.5, could be used for experimental verification of the above theoretical predictions. Furthermore, as an additional aspect, deep subwavelength nature of interactions, material granularity, and spatial dispersion could impact various types of newly proposed phenomena, e.g. self-induced torques [29] and cloaking [30], to name just a few. Emulation experiments, similar to the one, proposed here, could serve as auxiliary tools and enable further detailed investigations.

For the summary, numerical and experimental investigations of scattering from objects, embedded in a metamaterial assembly, were reported. In particular, tailoring electromagnetic properties of the embedding medium was demonstrated to provide a tool for significant shifting of the scattering resonance frequency due to a strong coupling between a scatterer and a metamaterial. Moreover, it has been shown that the direct electric contact between the scatterer and the embedding metamaterial dramatically changes the interaction nature, making it to be conductive instead of capacitive. Shortage between the wires causes the almost twice-stronger effect in terms of the resonance shift. As for a brief comparison, controllable investigation of contact/noncontact regimes between closely situated complex nanostructures is a very challenging task, underlining the strength of the emulation approach. Metamaterials, and wire medium in particular, could create the reduced space for modes, available for wave diffraction and, as the result, manipulate scattering properties of embedded objects. Creation of reduced diffraction and controlled scattering regimes is beneficial for various applications, e.g. invisibility cloaking [31]. Detailed emulation experiments with microwaves pave a way for designing experiments in the optical domain.

## Acknowledgments:

This work was partially supported by the Government of the Russian Federation (Grant 074-U01), by the Ministry of Education and Science of the Russian Federation (project 11.G34.31.0020, GOSZADANIE 2014/190, Zadanie No.3.561.2014/K, Project No. 14.Z50.31.0015), by Russian Foundation for Basic Research. The work on numerical simulations and investigation of the field distributions has been funded by the Russian Science Foundation Grant No. 14-12-01227. P.G. acknowledges the support from TAU Rector Grant and German-Israeli Foundation (GIF, grant number 2399).